\documentclass{phb-proc4-auth}

\usepackage{graphicx}
\usepackage{amssymb,amsmath}
\usepackage{subfigure}
\usepackage{cite}                     %

\begin{document}
\newcommand{\UA}{\uparrow}
\newcommand{\DA}{\downarrow}
\newcommand{\cdag}{c^\dagger}
\newcommand{\cnod}{c^{\phantom{\dagger}}}
\newcommand{\E}{{\textrm{e}}}

\begin{frontmatter}

\journal{SCES '04}
\date{June 25, 2003}


 \title{Single particle spectrum of the flux phase in the FM Kondo Model}


\author[TUG]{M. Daghofer\corauthref{1}}
\author[IC]{W. Koller}
\author[TUG]{W. von der Linden}
\author[TUG]{H. G. Evertz}


\address[TUG]{Institute for Theoretical and Computational Physics,
   Graz University of Technology, Petersgasse 16, A-8010 Graz,
   Austria}
\address[IC]{Department of Mathematics,
   Imperial College, 180 Queen's Gate, London SW7 2BZ, UK}

\corauth[1]{daghofer@itp.tu-graz.ac.at }


\begin{abstract}
We investigate the 2D ferromagnetic Kondo lattice model for
manganites with classical corespins 
at Hund's rule coupling $J_H=6$, with antiferromagnetic superexchange $0.03 \leq J'\leq 0.05$.
We employ canonical and grand canonical unbiased Monte Carlo simulations 
and find paramagnetism, weak ferromagnetism and the Flux phase,
depending on doping and on $J'$. 
The observed single particle spectrum in the flux phase differs from the idealized 
infinite lattice case,
but agrees well with an idealized finite lattice case with thermal fluctuations.
\end{abstract}



\end{frontmatter}

The ferromagnetic Kondo lattice model is believed to describe the electronic
degrees of freedom of compounds such as manganites. 
In these materials the
spins of the itinerant electrons have a strong ferromagnetic coupling to localized
corespins $\mathbf S_i$,
leading to
an Effective Spinless Fermion (ESF) Hamiltonian
\cite{KollerPruell2002a}
\begin{multline}                                      \label{eq:H}
  \hat H = -\sum_{<i,j>} t^{\UA\UA}_{i,j}\,
    \cdag_{i}\,\cnod_{j} - \sum_{i,j}
    \frac{t^{\UA\DA}_{i,j}\,t^{\DA\UA}_{j,i}}{2J_\textrm{H}}\, \cdag_{i}\cnod_{i}\\
    + J'\sum_{<i,j>} \mathbf S_i \cdot \mathbf S_j \;,
\end{multline}
where $\cdag_{i}$ creates an electron at site $i$ with its spin parallel to
$\mathbf S_i$, and $J_H=6$ throughout this work.
The hopping of these spinless electrons is modulated by the factor
$
  t^{\UA\UA}_{i,j}\;=\;\cos(\vartheta_{ij}/2)\;\E^{i\psi_{ij}},
$
which depends on the relative angles $\vartheta_{ij}$ of the two neighboring
corespins and carries a complex phase $\psi_{ij}$. On a two-dimensional
lattice, this complex phase leads to the so call ``Flux phase'' around half
filling of the lower Kondo band (doping $x=50\%$)~\cite{Flux_Litt}.
While the system is ferromagnetic at half filling for $J' \lesssim
0.025$, the Flux phase has been shown~\cite{Flux_Litt} 
to be the ground state at
half filling for larger values of $J'$. 

\begin{figure}[t]
  \includegraphics[width=0.45\textwidth]{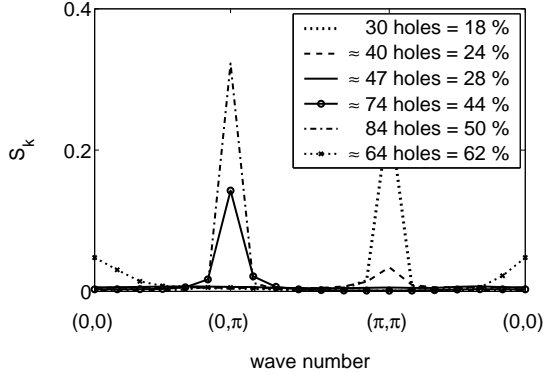}
  \caption{Corespin structure factor for $J_H=6, \beta=50$, $J'=0.05$ and
    various doping levels. Simulations were done on a $12 \times 14$ lattice,
    i.\,e. for $L=168$ lattice sites.\label{fig:ssk}} 
\end{figure}

We performed canonical and grand canonical unbiased Monte Carlo 
simulations~\cite{DaghoferKoller2003}.
The magnetic phases can be observed in the corespin structure factor
which is depicted in Fig.~\ref{fig:ssk} for $J'=0.05$.
In the polaronic regime with high compressibility 
at $x \approx 18\%$ holes~\cite{DaghoferKoller2003}, 
only an AFM signal is visible.
At $x\approx 24\%$, the compressibility is suddenly much reduced and the
system becomes homogeneous but a small AFM peak persists.
At $x\approx 28\%$ no magnetic structure at all is visible,
The flux phase appears at 
$x \approx 44\%$ and $x=50\%$ with signals at $(\pi, 0)$ and $(0,\pi)$.
The only FM signal exists, weakly, at even larger doping $x \approx 62\%$.

\begin{figure}[t]
  \includegraphics[width=0.45\textwidth]{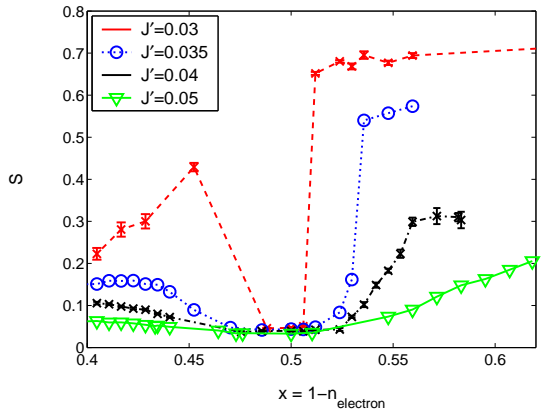}
  \caption{FM corespin Moment $S(q=0)$ as a function of doping for various values
    of $J'$, remaining parameters as Fig.~\ref{fig:ssk}.\label{fig:coresp}} 
\end{figure}

Fig.~\ref{fig:coresp}
shows the normalized FM corespin moment $S(q=0) = |\sum_i \vec S_i |/L$ as a function of
doping, where $L$ is the number of lattice sites.
It is expected to vanish in the Flux phase.
One sees that 
the doping range of the Flux phase becomes smaller 
at smaller $J'$.
The lattice is weakly ferromagnetic 
at slightly lower doping $x \approx 40\%$. 
All our data are compatible with the existence of 
a flux phase in the region where $S(q=0)\simeq 0$.


\begin{figure}[t]
  \centering
  \subfigure[]{
  \begin{minipage}{0.12\textwidth}
  \vspace*{3em}
  \includegraphics[width=\textwidth]{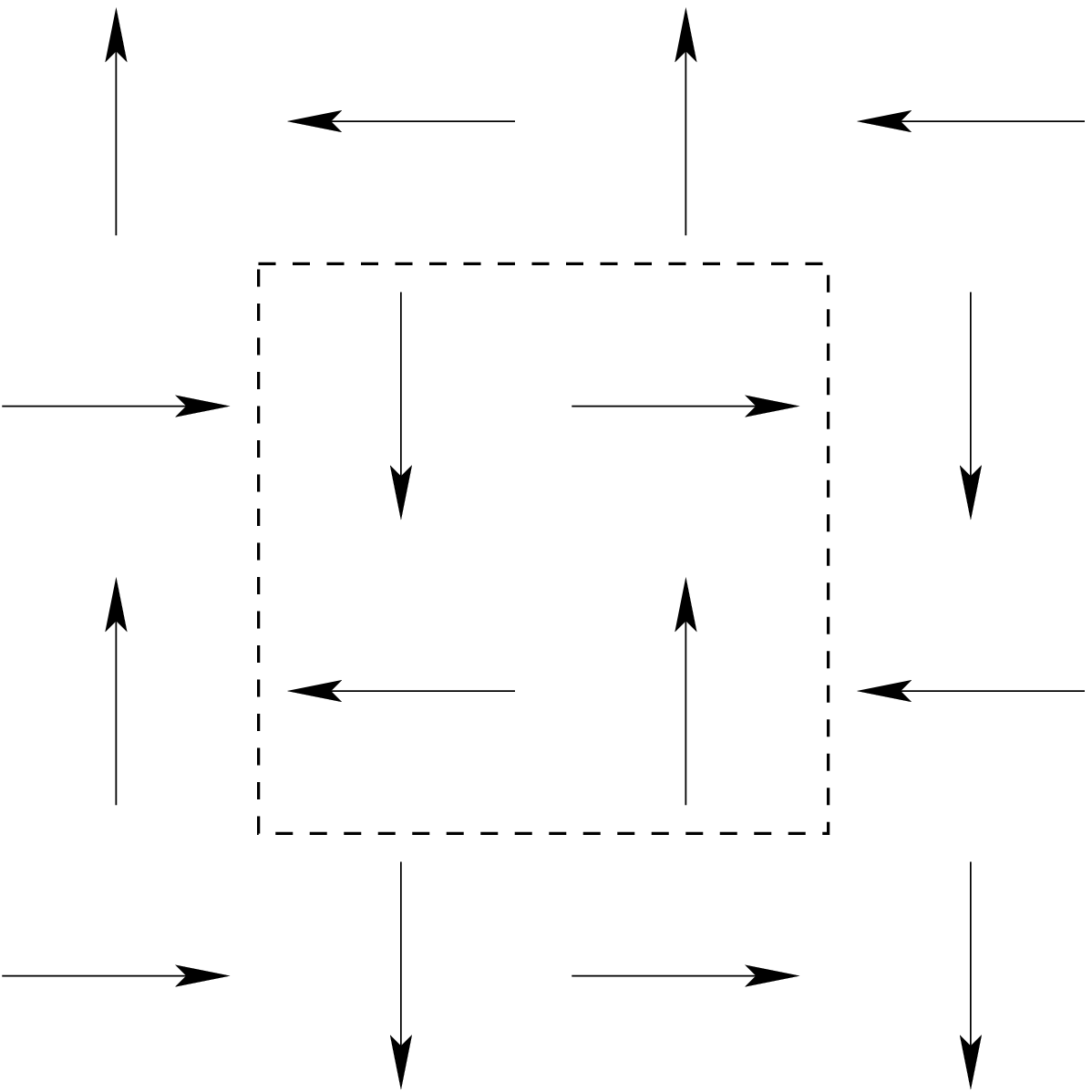}
  \vspace{1em}
  \end{minipage}\label{fig:flux_spins}}
  \hfill
  \subfigure[]{
  \begin{minipage}{0.3\textwidth}
  \includegraphics[width=\textwidth]{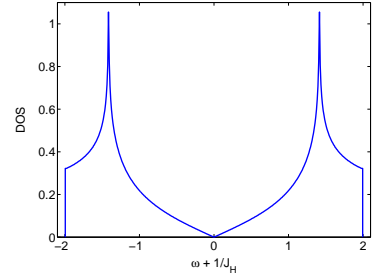}
  \end{minipage}\label{fig:flux_dos}}
  \caption{Perfect Flux phase: (a): Schematic plot of the corespin
 configuration.
 The dashed square delimits the unit cell. All spins are drawn parallel to
 the $xy$-plane.
 (b): Electronic DOS for the ESF Hamiltonian with
 $J_\textnormal{H} = 6$ on an infinite lattice.\label{fig:flux_schem}} 
\end{figure}

The corespin structure 
in an idealized  flux phase
is depicted in Fig.~\ref{fig:flux_spins}.
While all corespins are drawn parallel to the $xy$-plane for better
visibility, a global rotation in spin space leads to 
equivalent configurations.
In such corespin configurations, 
$|t^{\UA\UA}_{i,j}|=1/\sqrt 2$ is constant.
However, the hopping carries a complex phase~$\psi_{ij}$. 
Diagonalizing the ESF Hamiltonian for the four site unit cell of the Flux phase
leads to the dispersion relation
\begin{equation}                                    \label{eq:flux_esf_disp}
  \epsilon( k_x, k_y ) = -\tfrac{1}{J_\textnormal{H}} \pm \sqrt{2
    (\cos^2k_x + \cos^2k_y)}\;.
\end{equation}
The one-particle DOS of these bands for the infinite lattice with the
corespins aligned in the perfect Flux phase is depicted in
Fig.~\ref{fig:flux_dos}; it features a pseudogap at half filling.
The dispersion and the DOS of the full Kondo model and for
$J_\text{H}\rightarrow\infty$ can be found in \cite{Flux_Litt}.
\begin{figure}[t]
  \includegraphics[width=0.22\textwidth]{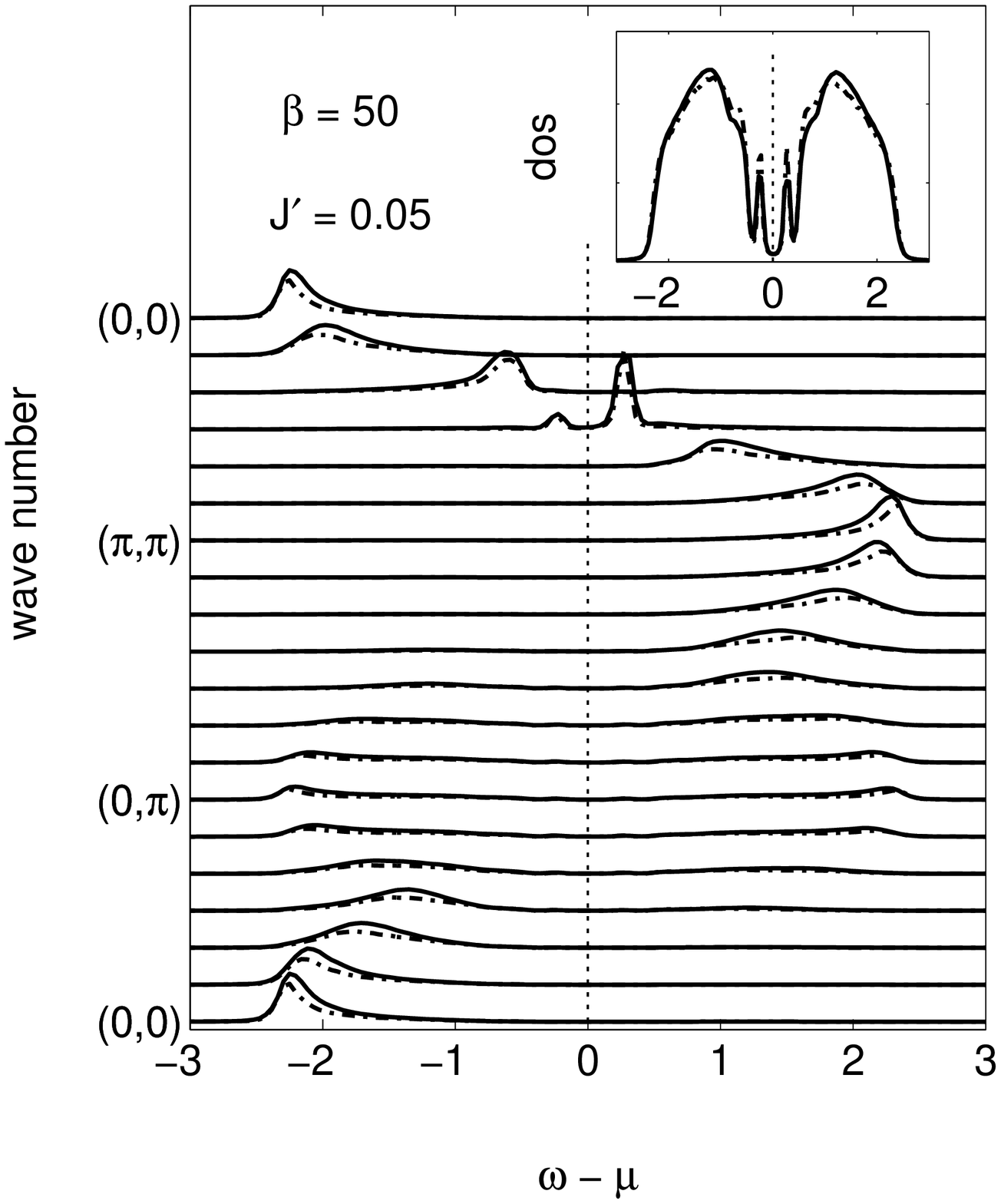}
  \hfill
  \includegraphics[width=0.22\textwidth]{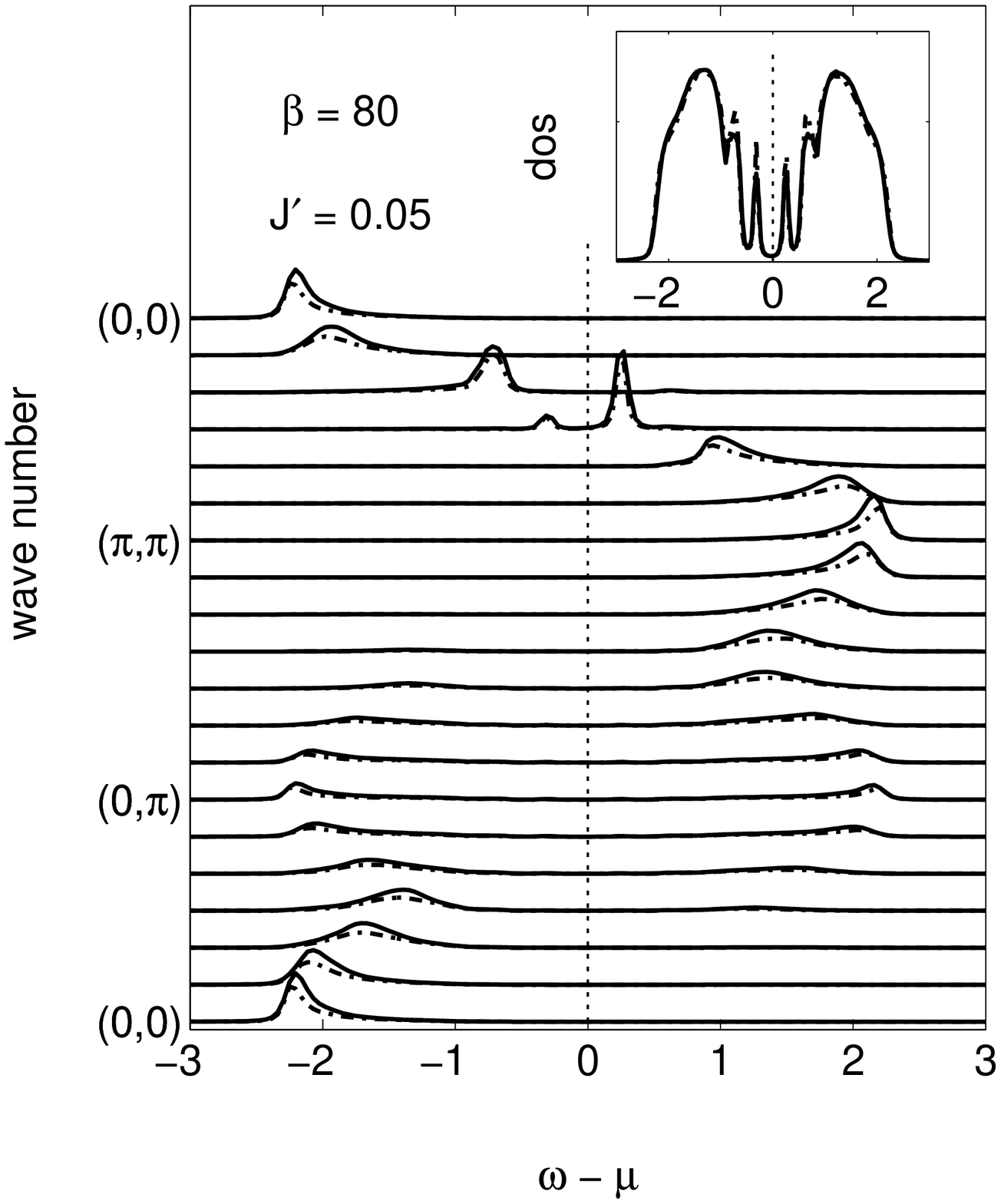}
  \caption{Spectral density for the Flux phase at half filling for $J' =
  0.05$; $\beta = 50$ (left) and $\beta = 80$ (right) on a $14 \times
  12$-lattice. Dashed lines: MC results for the ESF model. Solid lines: ideal
  Flux phase with random deviations.}
  \label{fig:dos_MC_Jse0.05vsFlux}
\end{figure}

The spectral density of the Flux phase is subject to strong finite
size effects, which are caused by the fact that only a very small part of the Brillouin zone
contributes to the few states around the pseudogap at $(\pi/2,\pi/2)$.
In order to ascertain that we did indeed observe the Flux phase, we therefore
compared our MC results to a model system. 
We started from a perfect Flux phase configuration 
of the corespins on a finite lattice, added random
perturbations, and diagonalized the resulting ESF Hamiltonian.
The amount of fluctuations was chosen to fit the width of the peaks in
the full MC results.
Fig.~\ref{fig:dos_MC_Jse0.05vsFlux} compares the average spectral density
and the DOS thus obtained to MC simulations for the ESF model
exhibiting almost perfect agreement,
while there is a notable difference between the DOS 
in Fig.~\ref{fig:dos_MC_Jse0.05vsFlux} and the idealized 
case in Fig.~\ref{fig:flux_dos},
The system can thus indeed be well described
by thermal fluctuations about the Flux phase on a finite lattice.

\textbf{Acknowledgments}:
We would like to thank the Austrian Science Fund (FWF project P15834)
and the EPSRC (Grant GR/S18571/01) for financial support.

\end{document}